\documentclass{article}

\PassOptionsToPackage{numbers,compress}{natbib}

\usepackage[preprint]{neurips_2026}

\usepackage[utf8]{inputenc}
\usepackage[T1]{fontenc}
\usepackage{amsmath, amssymb, amsthm}
\usepackage{graphicx}
\usepackage{booktabs}
\usepackage{multirow}
\usepackage{microtype}
\usepackage{xcolor}
\usepackage{hyperref}
\usepackage{url}
\usepackage{enumitem}
\usepackage{caption}
\usepackage{subcaption}
\hypersetup{
  colorlinks=true,
  linkcolor=blue!50!black,
  citecolor=blue!50!black,
  urlcolor=blue!60!black
}

\newcommand*\samethanks[1][\value{footnote}]{\footnotemark[#1]}

\title{The Fusion Equilibrium Challenge:\\ Inferring Magnetic Geometry Without Magnetic Diagnostics}

\author{
    Tapan Ganatma Nakkina\thanks{Sophelio, Austin, TX USA} \and 
    \textbf{Matthew Waller}\samethanks[1] \and 
    \textbf{Craig Michoski}\samethanks[1] \and
    \textbf{Brian Sammuli}\thanks{General Atomics, San Diego, CA, USA} \and
    David R. Hatch\thanks{University of Texas, Austin, TX, USA} \and
    William Boyes\samethanks[2] \and
    Mitchell Clark\samethanks[2] \and
    Raffi Nazikian\samethanks[2] \and
    Sterling Smith\samethanks[2] \\ \and  
{\tt michoski@sophel.io}}

\begin{document}
\maketitle

\begin{abstract}
Next-generation fusion reactor devices such as SPARC, ARC, and CFETR will operate in extreme neutron environments that compromise the magnetic sensors traditionally used to reconstruct plasma equilibria. However, reliable knowledge of the plasma equilibrium---including magnetic flux surfaces, safety factor profiles, and shaping parameters---is indispensable for real-time control, disruption avoidance, and physics interpretation. The \emph{Fusion Equilibrium Challenge} invites the NeurIPS community to confront a deceptively simple but scientifically rigorous inverse problem: \emph{reconstruct the two-dimensional poloidal flux function $\psi(R,Z)$ and a suite of scalar equilibrium parameters from non-magnetic diagnostics alone}, namely external poloidal-field coil currents and Thomson-scattering electron temperature/density profiles. 
The challenge provides the first open-access, harmonized multi-machine benchmark for fusion, releasing a curated dataset of 9{,}113 DIII-D shots and 2{,}416 MAST shots --- filtered for Thomson-diagnostic availability, feature completeness, and EFIT-reconstruction quality. Each shot is packaged into a standard Parquet file containing $\sim$260 (DIII-D) / $\sim$80 (MAST) EFIT flux maps and rich high-rate diagnostics. Two complementary awards reward intra-machine reconstruction fidelity ($S_{\text{model}}$) on DIII-D and zero-shot cross-machine generalization ($G_{\text{ratio}}$) to the topologically distinct MAST spherical tokamak. 
We argue that the challenge functions as a benchmark for reactor-ready equilibrium inference and as a probe of how far machine learning can be pushed toward truly machine-agnostic plasma state estimation.
\end{abstract}


\section{Competition Description}
\label{sec:competition}

\subsection{Background and Impact}
\label{sec:background_impact}

Magnetic confinement fusion is transitioning from physics-exploration tokamaks to a new generation of \emph{reactor-class} devices designed to demonstrate net energy gain, e.g., SPARC \citep{creely2020overview, kuang2018physics}, the high-field compact pilot plant ARC \citep{sorbom2015arc}, the Chinese demonstrator CFETR \citep{wan2017overview}, and ultimately ITER and DEMO \citep{romanelli2009overview}. These devices will deliver neutron fluxes more than an order of magnitude higher than any existing facility. In this environment, the very diagnostics that have underpinned tokamak control for half a century, i.e., Mirnov coils, pickup loops, and Rogowski coils mounted on or near the vessel walls, face severe degradation. Radiation-induced electromotive force (RIEMF) and radiation-induced conductivity (RIC) can produce drift voltages that exceed the signal of interest in long pulses, and structural transmutation reduces the sensor lifetime to a small fraction of the operational hours required of a commercial plant \citep{nishitani2014performance, vayakis2008status}.

\begin{figure}[h]
    \centering
    \includegraphics[width=0.95\textwidth]{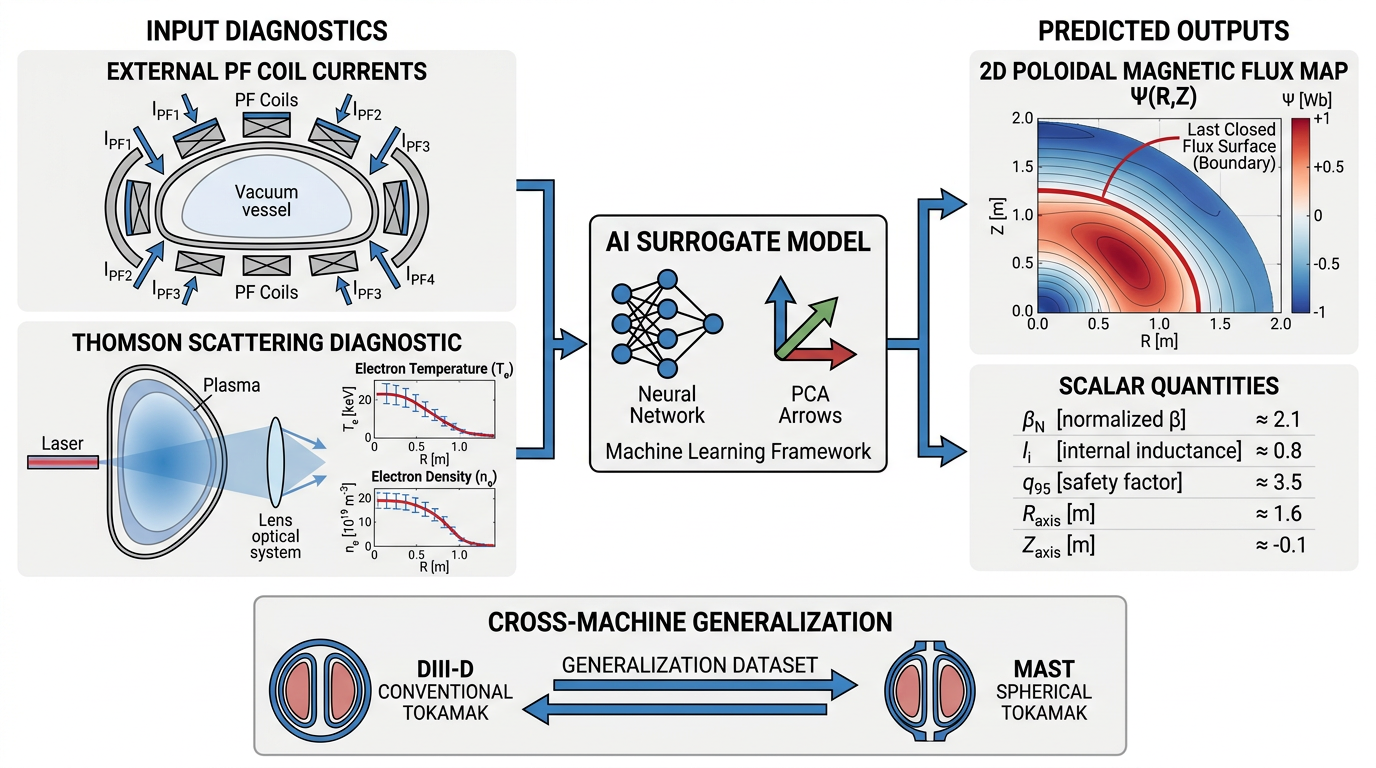}
    \caption{\textbf{Graphical abstract.} The Fusion Equilibrium Challenge asks participants to predict the two-dimensional poloidal flux map $\psi(R,Z)$ --- and with it the last closed flux surface (LCFS) and the scalar equilibrium parameters, which the scorer derives from the submitted flux map --- from non-magnetic diagnostics only: external coil currents and Thomson-scattering electron temperature and density profiles. Models are evaluated both on the DIII-D conventional tokamak and on the MAST spherical tokamak, with cross-machine transfer as a first-class objective.}
    \label{fig:graphical_abstract}
    \vspace{-15pt}
\end{figure}

The equilibrium-reconstruction problem itself does not disappear; it becomes more important. Plasma control loops need real-time, high-fidelity estimates of the poloidal magnetic flux $\psi(R,Z)$ to position the strike points, control the safety factor, avoid vertical displacement events, and prevent disruptions \citep{ferron1998real, piovesan2024integrated}. One popular numerical solver --- the equilibrium-fitting (EFIT) code of \citet{lao1985reconstruction}, refined over four decades \citep{lao2005mhd} 
--- weights an overdetermined least-squares fit of magnetic flux loops and pickup coils against a Grad--Shafranov solution \citep{grad1958hydromagnetic, shafranov1966plasma}. Strip away the magnetic signals and EFIT loses essentially all of its constraints: \emph{the very capability we will demand of next-generation devices is the one our existing toolkit cannot deliver}.

The challenge therefore poses a question that is both scientifically deep and operationally urgent: \emph{can machine learning, trained on present-day machines, learn an equilibrium representation that survives the loss of magnetic diagnostics?} A successful answer would (i) provide a backup estimator for present machines, useful when magnetic sensors fail or saturate; (ii) form the basis of a \emph{primary} equilibrium pipeline for reactor-class devices; and (iii) by explicitly demanding cross-machine generalization, test whether learned models can capture the universal physics of the Grad--Shafranov equation rather than the engineering minutiae of a single device. We anticipate strong engagement from the fusion-ML subcommunity \citep{degrave2022magnetic, seo2024avoiding, kim2024gsdeepnet, lao2022application, abbate2021data, wang2024efitnn}, the scientific-ML community studying physics-informed networks \citep{raissi2019physics}, neural operators \citep{li2021fourier}, and image-to-image regression.

\subsection{Novelty}
\label{sec:novelty}

To the best of our knowledge, this challenge constitutes the first public release of curated, multi-machine experimental fusion data under a permissive open license, transitioning tokamak equilibrium reconstruction from a closed-access institutional task to a standardized open-science benchmark.
Existing ML-for-fusion work falls into three categories: accelerated EFIT surrogates trained on full magnetic inputs \citep{joung2019deep, kim2024gsdeepnet, wang2024efitnn, lu2023fast, lao2022application}; downstream task models for disruption \citep{kates2019predicting, rea2019disruption, churchill2020deep, aymerich2022disruption, vega2022disruption, cannas2007support} and transport \citep{abbate2021data, vandeplassche2020fast, morosohk2021machine, rodriguez2022nonlinear}; and reinforcement learning for plasma shape control \citep{degrave2022magnetic, seo2024avoiding, tracey2024towards}. None of these deliberately removes the magnetic signal pathway, which is the entire premise of our challenge. The challenge introduces three additional novel design elements:

\textbf{(1) Open dataset with a built-in exploration tool.} All shot data are released as one-row-per-shot Parquet on Hugging Face, alongside the cross-platform \emph{Data Fusion Labeler} (\href{https://dfl.sophelio.io/}{dFL}) desktop app that lets participants visualize $\psi(R,Z)$, coil currents, and Thomson profiles interactively, lowering the entry barrier for newcomers without fusion expertise.

\textbf{(2) Topologically distinct cross-machine transfer.} DIII-D is a conventional tokamak with a D-shaped cross-section, while MAST/MAST-U is a low-aspect-ratio spherical tokamak whose plasma wraps tightly around a thin central post \citep{morgan2014mast, harrison2019overview}. The flux topologies are not merely scaled: DIII-D's contours form a concentric bullseye while MAST's form a kidney-bean wrapped around a hollow center column. A na\"ive coil-mapping transfer in our pilot collapses from SSIM~$0.83$ to $0.10$ --- effectively a failure, a more honest stress test than within-machine splits.

\textbf{(3) Consistency-based scoring that cannot be gamed by a scalar head.} Participants submit only the flux map plus the two scalars a flux map cannot contain ($q_{95}$ and $\beta_N$). Every other scored quantity --- the LCFS boundary, magnetic axis, shape, plasma volume and internal inductance --- is \emph{derived from the submitted flux map} by the scorer, with the same published functional applied to the ground-truth flux, so agreement is scored between $f(\hat\psi)$ and $f(\psi)$. A regression head decoupled from the predicted flux can no longer earn scalar points; a perfect flux map earns them all by construction. Teams may additionally label their data-preparation (\emph{harmonization}) pipeline in the submission manifest --- descriptive metadata that supports the post-competition analysis but is not scored.

\begin{figure}[t]
    \centering
    \includegraphics[width=0.82\textwidth]{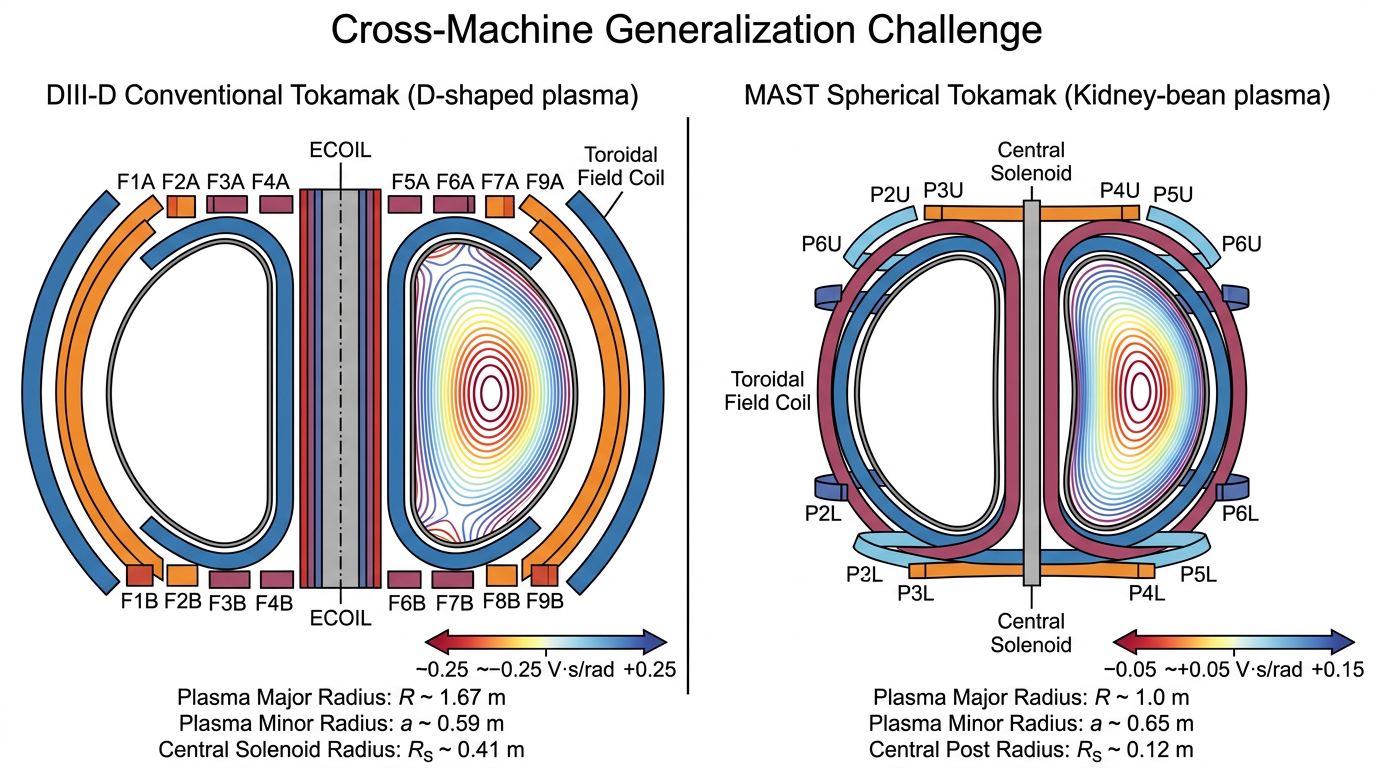}
    \caption{The two host devices for the challenge. \emph{Left:} DIII-D is a conventional tokamak with a large central solenoid (ECOILA), 18 shaping F-coils, and a D-shaped plasma cross-section. Flux values are negative ($\sim -0.25$~V$\cdot$s$/$rad). \emph{Right:} MAST is a spherical tokamak with a slender central column, ten P-coils, and a kidney-bean plasma profile; flux values are small and positive ($\sim +0.05$~V$\cdot$s$/$rad).}
    \label{fig:tokamak_comparison}
\end{figure}

\subsection{Data}
\label{sec:data}

\paragraph{Sources and scale.} The dataset is assembled from the DIII-D National Fusion Facility (General Atomics) \citep{strait2006magnetic, ponce2010thomson, eldon2017nonaxisymmetric} and the Mega Ampere Spherical Tokamak (MAST) at the Culham Centre for Fusion Energy \citep{morgan2014mast, harrison2019overview, pearson2010thomson}. The release contains 8{,}794 DIII-D shots and 2{,}414 MAST shots ($\sim$11{,}200 total), each stored as a single-row Parquet file with all multi-rate diagnostics nested as NumPy arrays. Records use machine-prefixed identifiers (e.g., \texttt{DIII-D\_182494}, \texttt{MAST\_25607}). Four filters were applied in sequence to arrive at this release: (i) the upstream host-laboratory ``good shot'' list (Thomson scattering available); (ii) on MAST, feature completeness (tangential ``edge'' Thomson present); (iii) on both machines, a minimum of 50 reconstructed EFIT slices per shot to exclude degenerate reconstructions; and (iv) the diverted-timeslice restriction described next. The dropped-shot manifests are published alongside the data as auditable companion files.

\paragraph{Diverted timeslices only.} The release contains \emph{only} diverted plasmas: limited frames, in which the last closed flux surface is set by physical contact with a material surface rather than by a magnetic x-point, are removed from \texttt{efit\_times} and from every column defined on it. The reason is that a limited boundary is fixed by the wall, and wall geometry is not contained in $\psi(R,Z)$; a boundary extracted from the flux map alone therefore systematically overshoots, and asking entrants to reconstruct a boundary that their inputs cannot determine is not a meaningful test. Frames are classified from each machine's best available evidence for the same physical criterion. On DIII-D we use the sign of EFIT's a-file \texttt{DSEP} separatrix--limiter clearance, which retains $87.9\%$ of frames. MAST publishes no such label --- and, contrary to a natural expectation, the IMAS field \texttt{equilibrium.time\_slice[:].boundary.type} is absent from the FAIR-MAST archive --- so we use EFIT++'s own indicator, namely that it reports x-point coordinates only on frames where it located one, retaining $66.4\%$ of frames. We validate that indicator against the shipped flux itself by testing whether the x-point lies on the boundary flux surface, $|\psi(\mathbf{x}) - \psi_{\text{bdry}}| / |\psi_{\text{bdry}} - \psi_{\text{axis}}|$, which agrees on $97.3\%$ of $184{,}396$ frames with a median normalized offset of $6\times10^{-4}$. The full-rate diagnostic inputs are not subset: entrants receive the complete discharge and resample onto \texttt{efit\_times} as before.

\paragraph{Machine geometry as an input.} Because equilibrium reconstruction is a boundary-value problem, the release ships each machine's fixed conductor and diagnostic geometry alongside the time series: poloidal-field coil positions, cross-sections and turn counts, keyed to the current channels they belong to, and Thomson-scattering chord positions in both $R$ and $Z$. The two machines are described at their native granularity --- DIII-D as $19$ lumped rectangles with turn counts, matching EFIT's own \texttt{mhdin.dat} representation, and MAST as $812$ individual conductor elements from the FAIR level-2 \texttt{pf\_active} IDS. DIII-D chord positions are shipped per shot because they are not constant across campaigns ($11$ distinct subsystem layouts, $59$--$138$ channels).

\paragraph{Schema.} Each shot is stored as a single-row Parquet record whose columns contain either scalar metadata or nested NumPy arrays. DIII-D records include the EFIT targets (\texttt{efit\_times}, \texttt{efit\_psirz}), machine-specific magnetics streams for 18 shaping F-coils together with \texttt{ECOILA}, \texttt{bcoil}, plasma current, and the EFIT-derived \texttt{dsep} signal, plus Thomson-scattering core and edge measurements (44 and 10 spatial channels, respectively) and the coil/chord geometry described above. MAST records contain the same EFIT targets together with physical grid coordinates (\texttt{efit\_grid\_R/Z}), a shared magnetics time base, ten P-coils, central solenoid, toroidal-field coil, error-field-protection-system coil, plasma current, \texttt{dsep}, its own coil/chord geometry, and analogous core/edge Thomson measurements. The \texttt{dsep} columns are retained as context on both machines but are not scored, and they are \emph{not} the same physical quantity across the two (Section~\ref{sec:metrics}). This schema preserves each machine's native diagnostic structure while exposing a common per-shot storage format.

\paragraph{Multi-rate heterogeneity and NaNs.} Magnetics are recorded at $\sim$49{,}152 samples per shot on DIII-D and $\sim$15{,}482 on MAST, while EFIT flux maps appear only at $\sim$313 (DIII-D) and $\sim$77 (MAST) times per shot; Thomson systems operate on their own, machine-dependent intermediate cadences. Aligning these time bases \emph{without} corrupting the targets is the central data-engineering challenge --- the data-preparation (harmonization) work every pipeline must solve before any model sees a tensor. Raw MAST flux maps retain the native $65 \times 129$ grid, with roughly half of the entries marked NaN because they correspond to the physical central post rather than plasma-accessible space. For common-shape modeling and visualization, the benchmark materials also expose the corresponding valid-column $65 \times 65$ view.

\paragraph{Licensing and ethics compliance.} All shots are authorized by the host laboratories for public release and are distributed on Hugging Face under CC BY 4.0. The dataset complies with the NeurIPS Code of Ethics: it contains no human-subject data and no personally identifying information; the underlying experiments were performed by the host laboratories under their standard institutional safety protocols; the ground-truth EFIT reconstructions for the private test fold have not been previously published and remain confidential until the end of Phase~2; and feature-leakage risks are mitigated by enforcing strict shot-level train/test splits and by the reproducibility requirement on prize candidates (their pipeline must rebuild their submitted predictions from handed-over code and a pinned environment).

\begin{figure}[t]
    \centering
    \begin{subfigure}{0.49\textwidth}
        \centering
        \includegraphics[width=\textwidth]{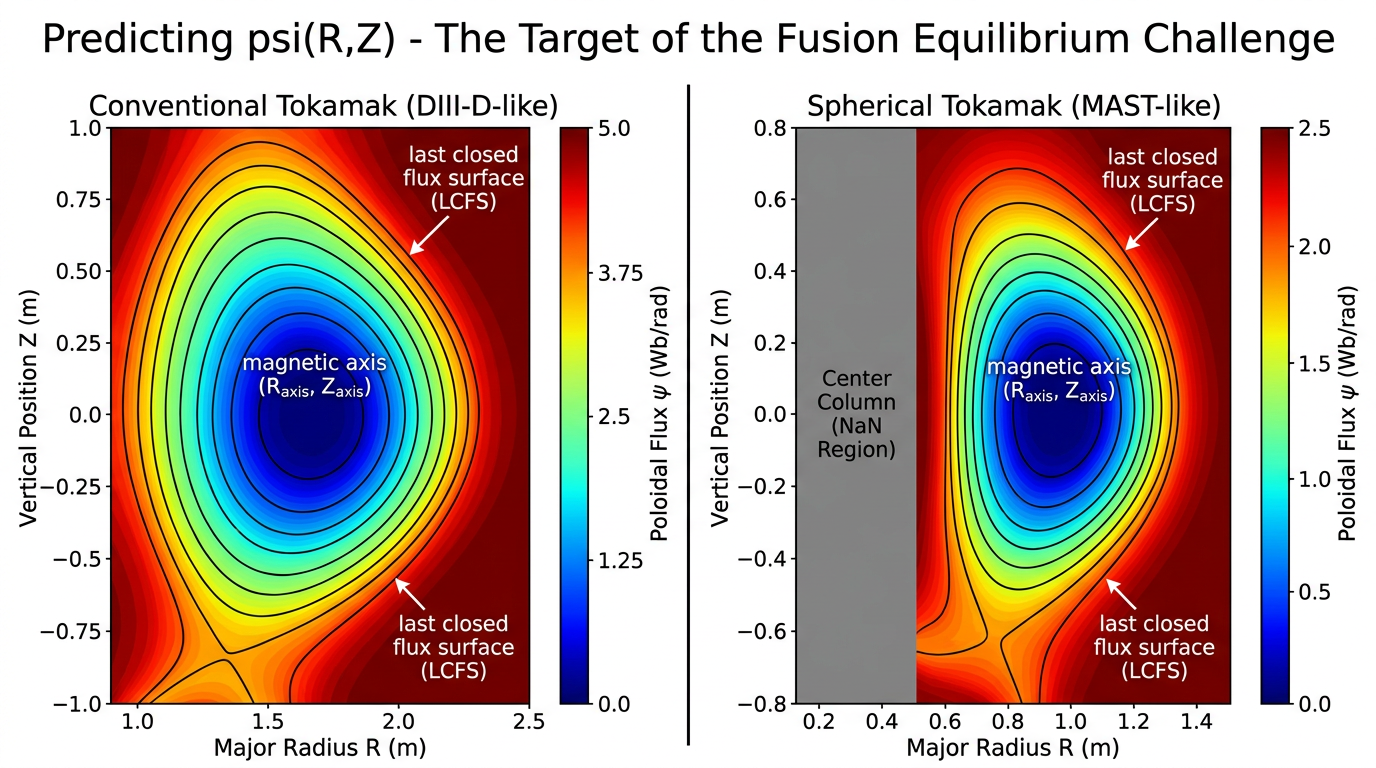}
        \caption{Illustrative $\psi(R,Z)$ targets for DIII-D (D-shape) and MAST (kidney bean, with gray NaN block in the central post).}
        \label{fig:flux_map_example}
    \end{subfigure}\hfill
    \begin{subfigure}{0.49\textwidth}
        \centering
        \includegraphics[width=\textwidth]{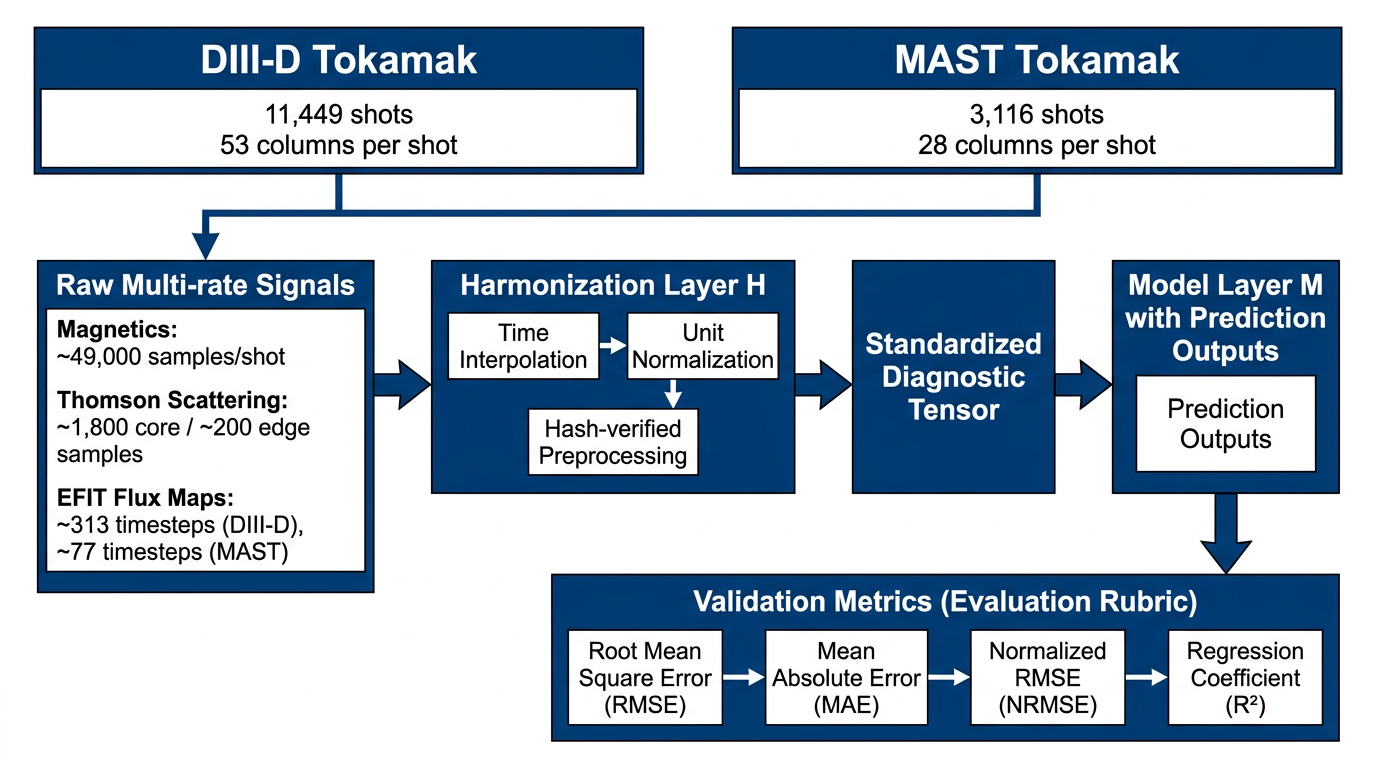}
        \caption{End-to-end data pipeline: raw multi-rate signals~$\rightarrow$~data-preparation (harmonization) pipeline~$\rightarrow$~model~$\rightarrow$~$(\psi, \text{LCFS}, \text{scalars})$.}
        \label{fig:data_pipeline}
    \end{subfigure}
    \caption{\emph{Left:} The 2D target. The challenge requires recovering nested flux surfaces, the LCFS contour, and the scalar equilibrium parameters from non-magnetic inputs only. \emph{Right:} The full data pipeline from raw multi-rate diagnostics through the participant's data preparation and model to the scored equilibrium.}
    \label{fig:flux_and_pipeline}
\end{figure}

\subsection{Tasks and Application Scenarios}
\label{sec:tasks}

\paragraph{Primary task.} Given the non-magnetic diagnostic inputs at all EFIT timestamps of a held-out shot, predict the full $65 \times 65$ poloidal flux map $\hat\psi(R,Z)$ together with the two scalars a flux map cannot contain --- the edge safety factor $q_{95}$ (which requires the toroidal field function $F(\psi)$) and normalized beta $\beta_N$ (which requires the pressure profile $p(\psi)$). Everything else the control room cares about --- the last closed flux surface (LCFS) contour, the magnetic-axis coordinates $(R_{\text{axis}}, Z_{\text{axis}})$, elongation $\kappa$, upper/lower triangularity $\delta$, plasma volume $V$, and internal inductance $l_i$ --- is \emph{derived from the submitted flux map} by the evaluation code and scored for consistency with the same derivation applied to the ground-truth flux (Section~\ref{sec:metrics}). These are the same quantities reported by traditional EFIT \citep{lao2005mhd} and form the canonical kernel of plasma-state information consumed by integrated control systems.

\paragraph{Compute-light badge.} A \emph{compute-light} badge recognizes submissions whose models train end-to-end in less than two hours on a single GPU or CPU (self-declared via the submission fact sheet); the baseline ridge regression and small CNN (Section~\ref{sec:baselines}) fit comfortably in this envelope. There is no separate track or leaderboard: all entries share the same data, submission format, and board, and unconstrained transformer-scale, diffusion-based, or neural-operator architectures compete alongside badge holders. The badge is a recognition rather than a quarantine, encouraging accessible solutions for educators and underprivileged participants.

\paragraph{Two awards.} The competition uses two complementary primary awards plus four \emph{honorable mentions}:
\begin{itemize}[topsep=2pt, itemsep=2pt]
    \item \textbf{Award \#1: Best DIII-D Intra-Machine Reconstruction (USD~500).} Highest composite $S_{\text{model}}$ on a hidden DIII-D test set.
    \item \textbf{Award \#2: Best DIII-D~$\rightarrow$~MAST Cross-Machine Generalization (USD~500).} Highest cross-machine generalization ratio $G_{\text{ratio}}$ among entries whose primary-domain $R^2_\psi > 0.6$.
    \item \textbf{Honorable Mentions.} Top performance on flux fidelity $R^2_\psi$, predicted-scalar fidelity $R^2_{q_{95},\beta_N}$, flux-map consistency (the $\psi$-derived scalar agreement of Section~\ref{sec:metrics}), and best LCFS alignment $D_{\text{LCFS}}$.
\end{itemize}

\paragraph{Application scenarios.} Several reactor-relevant operational scenarios motivate this task. (i) \emph{Diagnostic-loss redundancy}: in a present-day tokamak that experiences a transient magnetic-sensor failure --- a common cause of plasma-control loss --- a backup ML reconstruction could provide a safe shutdown trajectory. (ii) \emph{Reactor primary inference}: in SPARC and successor devices, kinetic diagnostics may outlast magnetic diagnostics, so an inference path that maps coil currents and electron profiles to $\psi(R,Z)$ becomes the dominant pipeline. (iii) \emph{Real-time control}: present EFIT runs at $\sim$10--100~ms latencies, which is too slow for fast modes (e.g., neoclassical tearing modes); a learned surrogate operating on the same input mix could plausibly run at sub-millisecond inference \citep{wang2024efitnn, kim2024gsdeepnet}. (iv) \emph{Physics interpretation}: kinetic-equilibrium reconstruction \citep{lao2005mhd} already integrates Thomson-scattering and magnetic data; our challenge effectively explores the limit where the magnetic terms are removed.

\paragraph{Societal impact and ethics compliance.} The task complies with the NeurIPS Code of Ethics. Its societal upside is to reduce the diagnostic burden on reactor-class fusion devices and thereby to support clean-energy R\&D; the technical content has no adversarial-use surface, since equilibrium reconstruction does not enable any new offensive capability. The only foreseeable harm is over-confidence in an ML reconstruction if it were operationally deployed without uncertainty quantification; we mitigate this by requiring participants to report bootstrap confidence intervals alongside point predictions and by framing the challenge explicitly as a benchmark rather than a deployment-ready system.

\subsection{Evaluation Metrics}
\label{sec:metrics}

The composite model score blends four terms, each with a diagnostic per-metric breakdown. The design principle is that a submission is a flux map plus only the two scalars a flux map cannot contain; every geometric quantity is computed \emph{from} the submitted flux with the same code that computes it from the ground truth, so it can only be earned by a flux map that implies it.

\paragraph{Flux reconstruction accuracy.} The pooled coefficient of determination,
\begin{equation}
    R^2_\psi \;=\; 1 \;-\; \frac{\sum_k \left(\psi_k - \hat\psi_k\right)^2}{\sum_k \left(\psi_k - \bar\psi\right)^2},
\end{equation}
where the index $k$ runs over all $(R, Z)$ grid points, all EFIT timesteps, and all shots in the test set; $\psi_k$ and $\hat\psi_k$ are the corresponding ground-truth and predicted fluxes; and $\bar\psi = \tfrac{1}{N}\sum_k \psi_k$ is the mean of the ground-truth flux over that same index set. We additionally report a per-shot mean of $R^2_\psi$ as a robustness diagnostic.

\paragraph{Last-closed-flux-surface alignment.} A length-normalized symmetric Hausdorff distance,
\begin{equation}
    D_{\text{LCFS}} \;=\; \frac{d_{\text{Haus}}(C, \hat C)}{\langle R_{\text{LCFS,true}}\rangle},
\end{equation}
where $d_{\text{Haus}}(\cdot,\cdot)$ is the symmetric Hausdorff distance between the predicted and true LCFS contours $C, \hat C$, and the denominator is the time-averaged true LCFS major radius. Smaller values are better. Participants do not submit a contour: the scorer extracts the LCFS from the predicted flux map with the same published procedure it applies to the ground-truth flux map, so a perfect flux map yields $D_{\text{LCFS}} = 0$ by construction.

\paragraph{Predicted-scalar fidelity.} For the two submitted scalars $q \in \{q_{95}, \beta_N\}$ --- the only equilibrium parameters not recoverable from $\psi(R,Z)$ alone, since $q_{95}$ requires the toroidal field function $F(\psi)$ and $\beta_N$ the pressure profile $p(\psi)$ --- we compute the pooled coefficient of determination against the stored EFIT values and average:
\begin{equation}
    R^2_{q_{95},\beta_N} \;=\; \tfrac{1}{2}\bigl(R^2_{q_{95}} + R^2_{\beta_N}\bigr).
\end{equation}

\paragraph{Flux-map consistency.} For each of the seven $\psi$-derived scalars
$j \in \{R_{\text{axis}}, Z_{\text{axis}}, \kappa, \delta_{\text{top}}, \delta_{\text{bot}}, V, l_i\}$
the evaluation code applies the same functional $f_j$ --- the standard EFIT-literature derivation
(O-point location for the axis; LCFS bounding geometry for $\kappa$/$\delta$; $V = 2\pi\iint R\,dR\,dZ$
inside the LCFS; $l_i$ in the ITER li(2) normalization via $B_p = |\nabla\psi|/R$ and Amp\`ere's law)
--- to both the predicted and the ground-truth flux map, and scores the pooled agreement
\begin{equation}
    R^2_{\text{cons},j} \;=\; 1 - \frac{\sum_k \bigl(f_j(\hat\psi_k) - f_j(\psi_k)\bigr)^2}{\sum_k \bigl(f_j(\psi_k) - \overline{f_j(\psi)}\bigr)^2},
    \qquad
    \text{Consistency} \;=\; \frac{1}{|J|}\sum_{j \in J} \max\bigl(0, R^2_{\text{cons},j}\bigr).
\end{equation}
Because $f_j$ is applied identically to both sides, unit and normalization conventions cancel, and a
perfect flux map scores $\text{Consistency} = 1$ exactly. Because the release is restricted to
\emph{diverted} timeslices (Section~\ref{sec:data}), every scalar is pooled over every frame:
$R^2_\psi$, $R^2_{q_{95},\beta_N}$ and Consistency share a single frame population, and on the MAST
public fold the reference LCFS extraction succeeds on $69{,}075/69{,}075$ frames. A derivation that
fails on the \emph{predicted} map is penalized by mean-substitution rather than skipped. Per-scalar
breakdowns and failure rates are reported with every submission.

We do not score the x-point gap \texttt{dsep}. Although both machines ship a column of that name,
they are not the same physical quantity: on DIII-D it is EFIT's a-file separatrix--limiter
\emph{clearance}, whose sign encodes limited versus diverted operation, whereas on MAST it is the
upper--lower divertor \emph{balance} $\delta R_{\text{sep}}$, which straddles zero for ordinary
diverted plasmas. A single functional cannot score both coherently, and with the release restricted
to diverted frames the classification role the signal also played is no longer required.

\paragraph{Clipping convention.} For the purpose of building the composite intra-machine score below, we clip the per-metric quantities to bounded ranges before aggregation: $R^2_\psi, R^2_{q_{95},\beta_N} \leftarrow \max(0, \cdot)$ (a constant predictor receives a score of $0$ rather than a large negative value) and $D_{\text{LCFS}} \leftarrow \min(1, D_{\text{LCFS}})$ (an LCFS error larger than the mean major radius is no worse than ``boundary completely missed''); the Consistency term is clipped per scalar as above. Unclipped values are reported separately for transparency.

\paragraph{Composite intra-machine score.} With clipped per-metric values,
\begin{equation}
    S_{\text{model}} \;=\; w_1\, R^2_\psi \;+\; w_2\, R^2_{q_{95},\beta_N} \;+\; w_3 \bigl(1 - D_{\text{LCFS}}\bigr) \;+\; w_4\, \text{Consistency},
\end{equation}
with organizer-default weights $(w_1, w_2, w_3, w_4) = (0.55, 0.15, 0.10, 0.20)$ summing to $1$, so $S_{\text{model}} \in [0, 1]$. The flux map remains dominant; the only genuinely independent scalars carry a real weight of their own; and the boundary and interior geometry terms together carry $0.30$, all of it earnable only through the submitted flux. The highest $S_{\text{model}}$ wins Award~\#1.

\paragraph{Generalization ratio.} The cross-machine transfer score is
\begin{equation}
    G_{\text{ratio}} \;=\; \frac{S_{\text{model}}^{\text{MAST}}}{S_{\text{model}}^{\text{DIII-D}}},
\end{equation}
i.e., how much of the primary-domain (DIII-D) performance the same end-to-end pipeline retains when applied zero-shot to MAST. The Award~\#2 admissibility gate ($R^2_\psi > 0.6$ on DIII-D) together with the clipping convention above guarantees $S_{\text{model}}^{\text{DIII-D}} > w_1 \cdot 0.6 = 0.33$. Since $S_{\text{model}} \in [0,1]$, it follows that $0 \leq G_{\text{ratio}} < 1/0.33 \approx 3.0$, so the ratio is well-defined and bounded. Values near $1$ indicate near-complete retention of source-domain performance, while values below $1$ indicate degradation under transfer. As an early indication, the pilot coil-mapping baseline shows a comparable collapse (an SSIM-based proxy for $G_{\text{ratio}}$ falls to $\approx 0.12$, i.e.\ $0.10/0.83$); the formal $S_{\text{model}}$-based $G_{\text{ratio}}$ is tracked on the public leaderboard.

\begin{figure}[t]
    \centering
    \includegraphics[width=0.82\textwidth]{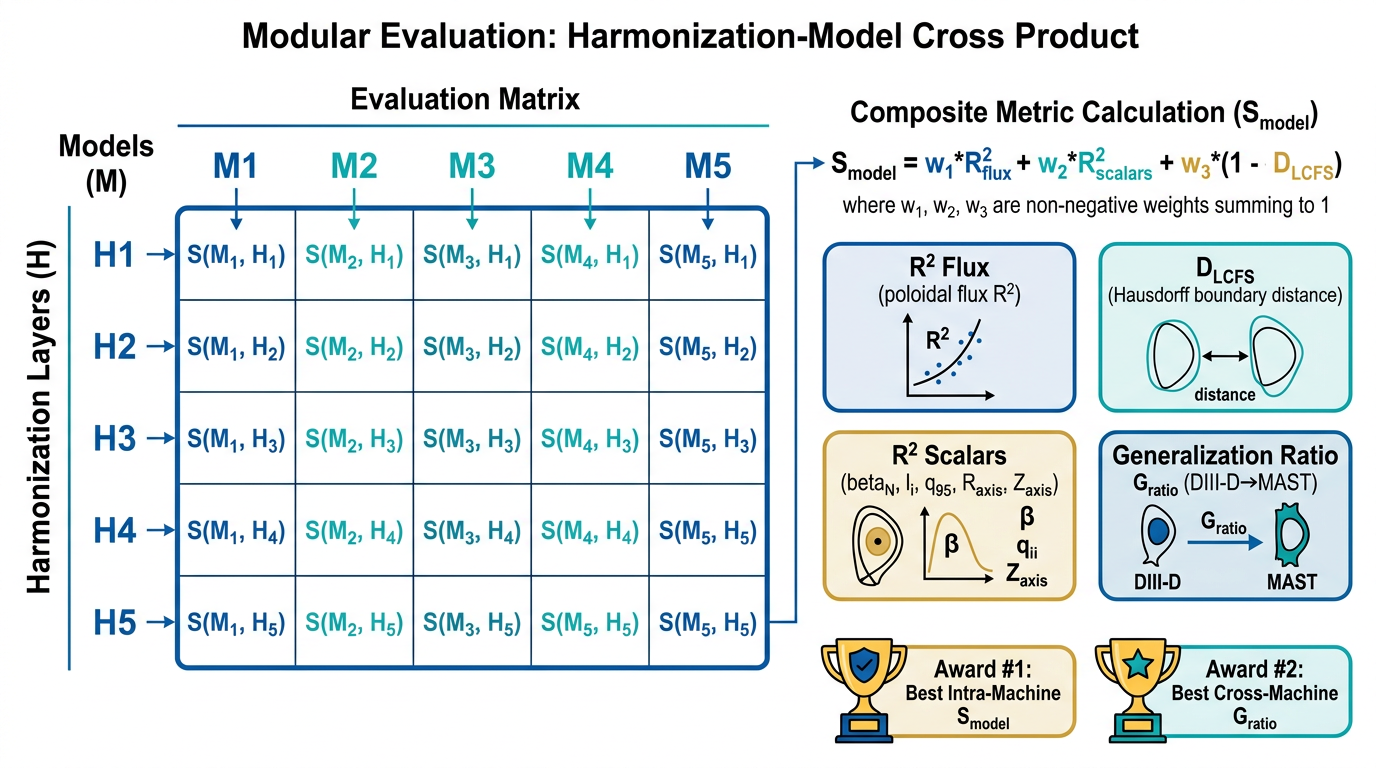}
    \caption{The composite intra-machine score $S_{\text{model}}$ weights flux fidelity, the two predicted scalars, LCFS alignment, and the consistency of the $\psi$-derived scalars --- the last three all computed from the submitted flux map itself; the generalization ratio $G_{\text{ratio}}$ measures the fraction of DIII-D performance retained on MAST.}
    \label{fig:evaluation_metrics}
\end{figure}

\subsection{Baselines, Code, and Material Provided}
\label{sec:baselines}

We release four reference baselines occupying different points on the complexity / data-efficiency frontier. All are implemented in PyTorch \citep{paszke2019pytorch} and scikit-learn \citep{pedregosa2011scikit} and released as Jupyter notebooks alongside the dataset. The pilot SSIM and MSE numbers reported below are quick visual-similarity proxies meant to give intuition about reconstruction quality; the formal leaderboard score is always the composite $S_{\text{model}}$ of Section~\ref{sec:metrics}, and its constituents $R^2_\psi$, $R^2_{q_{95},\beta_N}$, $D_{\text{LCFS}}$, and Consistency are reported per baseline on the public leaderboard.

\paragraph{Baseline 1 -- PCA + Ridge.} A 20-component PCA of training flux maps retains $\sim$99\% of the variance; \texttt{RidgeCV} fits the 21-dimensional DIII-D input vector to the 20 PCA coefficients. On a 3-shot demo set ($\sim$264 training samples) it achieves SSIM~$=0.84$ in $<\!0.1$ seconds. Linearity is justified because the magnetic field is approximately linear in the coil currents \citep{wesson2011tokamaks, freidberg2014ideal}.

\paragraph{Baseline 2 -- PCA + Gradient-boosted trees.} \texttt{HistGradient\-BoostingRegressor} in a \texttt{MultiOutput\-Regressor} captures leading nonlinear coupling between coils; interpretable via SHAP attributions.

\paragraph{Baseline 3 -- MLP on PCA targets.} A 3-layer fully-connected network (128 units, ReLU, dropout 0.1, Adam \citep{kingma2015adam}) predicting the same 20 PCA coefficients. With $\sim$41{,}000 parameters it achieves MSE~$=0.005$ on the demo set --- two orders of magnitude smaller than a pixel-wise alternative at comparable accuracy.

\paragraph{Baseline 4 -- Convolutional decoder.} A fully-connected encoder maps the input vector to a $4 \times 4 \times 32$ tensor, followed by four \texttt{ConvTranspose2d} layers upsampling to $65 \times 65$. With $\sim$5M parameters it overfits the demo set (MSE~$=0.239$) but surpasses PCA baselines on the full $10{,}000{+}$-shot training split, illustrating the capacity-vs-data tradeoff.

\paragraph{Cross-machine and synthetic-diagnostic baselines.} A na\"ive coil-mapping transfer (Baseline 1, alphabetical F\,$\rightarrow$\,P pairing) achieves SSIM~$=0.83$ on DIII-D but only $0.10$ on MAST. We additionally release a \emph{synthetic-diagnostic} starting kit that converts inputs into machine-agnostic physics features (electron pressure $p_e = n_e T_e$, stored-energy proxy $\sum n_e T_e$, $I_p$, $q \sim B_t/B_p$, up-down asymmetry).

\paragraph{Evaluation discipline and platform.} All baselines and the scoring code enforce \emph{shot-level} train/test splits (random timestep splits are a pernicious leakage pattern, since timesteps within a shot are highly correlated). All code is MIT-licensed (\url{https://github.com/fusion-equilibrium-challenge}); the dataset is hosted on Hugging Face; submissions are processed through Codabench \citep{rousseau2024fair} with public/private leaderboards, in \texttt{.npz} or NetCDF4 format indexed by sample IDs.

\subsection{Website, Tutorial, and Documentation}
\label{sec:website}

The challenge website at \texttt{fusion-equilibrium.challenge.io} hosts the motivation and timeline, a ``Getting Started'' walkthrough that guides newcomers through the ``jelly donut'' physical intuition and a first end-to-end submission in under fifteen minutes, leaderboards with bootstrap-uncertainty scores, a FAQ/forum on the Hugging Face Community tab and a GitHub Discussions board, and the cross-platform dFL desktop visualizer for $\psi(R,Z)$, coil currents, and Thomson profiles. Tutorial Jupyter notebooks reproduce the four baselines end-to-end with an estimated learning curve of 1--2 hours from setup to first submission.

\section{Organizational Aspects}
\label{sec:organization}

\subsection{Protocol}
\label{sec:protocol}

\paragraph{Steps to join.} Participants register a team account on the Codabench competition page, accept the contest rules and the dataset license terms, and pull the full Parquet dataset and the dFL visualizer directly from the Hugging Face repository (or one of the laboratory mirrors) using the official \texttt{datasets} library. No DIII-D or MAST credentials are required.

\paragraph{What participants submit.} A submission is a single archive uploaded to Codabench containing (i) the model's predictions for the public test shots as one \texttt{.npz} per machine, keyed by shot and EFIT timestamp --- the flux map $\hat\psi$ plus the $q_{95}$ and $\beta_N$ series, nothing else --- and (ii) a small manifest, which may carry an optional label naming the team's data-preparation (harmonization) pipeline; the label is descriptive metadata and is not scored. Final prize candidates additionally submit their full pipeline source code, a pinned environment specification, and a 1--2 page methods report so that organizers can independently rebuild and re-score the entry.

\paragraph{Phases and evaluation.} Phase~1 (\emph{development}, $\sim$3~months) opens with the training and validation Parquet files, baselines, and dFL visualizer; entries are scored on 50\% of the held-out shots and shown on a public leaderboard with bootstrap uncertainty estimates. Phase~2 (\emph{final}, one week) opens the blind half of the test set; the highest-scoring entry per team is forwarded for evaluation on this private fold, which determines the winners.

\paragraph{Preventing cheating and overfitting.} Submissions are capped at 5 per day and 100 in total to discourage leaderboard probing; all train/test splits are strictly shot-level; the consistency-based metric removes the scalar-head gaming surface (every geometric scalar must be earned by the submitted flux map); prize candidates' pipelines must reproduce their submitted predictions from the code and pinned environment they hand over; and the ground-truth EFIT reconstructions for the private fold are never released. The organizers may disqualify entries that exploit ground-truth leakage or test-set memorization.

\paragraph{Fairness for less well-resourced teams.} Every reference baseline trains to within a few percent of the leading published numbers in under two hours on a single commodity GPU or CPU, so participants without dedicated accelerator allocations can submit competitive entries. The compute-light recognition (Section~\ref{sec:tasks}) makes this explicit on the leaderboard.

\subsection{Rules and Engagement}
\label{sec:rules}

\paragraph{Verbatim rules.}
\begin{enumerate}[topsep=2pt, itemsep=2pt, leftmargin=*]
    \item Each team registers on Codabench with a single valid contact email; aliases that already exist as solo participants are deactivated.
    \item External public datasets (other tokamak archives, OMFIT-produced equilibrium tables) and publicly available pre-trained vision or scientific foundation models are permitted with explicit disclosure in the methods report.
    \item No restriction is placed on programming language or framework.
    \item All train/test splits must respect shot-level boundaries.
    \item Harmonization layers must regenerate deterministically under a clean rebuild (SHA-256 hash match).
    \item Top three teams in each award category must release their source code under an OSI-approved licence and submit a 1--2 page methods report before prizes are paid.
    \item Organizing-team members with access to the hidden ground truth are excluded from prize eligibility.
    \item All participants follow the NeurIPS Code of Conduct.
\end{enumerate}

\paragraph{How these rules support the desired outcome.} Shot-level splits and the deterministic-rebuild requirement protect both the leaderboard and the cross-machine generalization ranking from the most common leakage failures. Free use of language, framework, and pre-trained models keeps the entry barrier low, while the methods-report and code-release obligations make every winning solution reproducible and reusable by the fusion-ML community. Explicit disclosure of external data sources, and the exclusion of organizers from prize eligibility, avoid conflicts of interest without restricting participation. None of the rules penalises teams that fail to make a complete submission---unsuccessful entries simply do not appear on the leaderboard.

\paragraph{Communication with organizers.} Participants reach the organizers through three complementary channels: a dedicated email alias (\texttt{organizers@fusion-equilibrium.challenge.io}), the Hugging Face Community tab on the dataset page for dataset-specific discussion, and GitHub Discussions/Issues on the starter-kit repository for code and platform questions. Rule clarifications, deadline updates, and platform announcements are posted on the challenge website, on Codabench, and broadcast via a low-volume mailing list to which all registered participants are subscribed.

\subsection{Schedule and Readiness}
\label{sec:schedule}

\begin{table}[h]
\centering
\caption{Tentative competition timeline. Items marked $[\checkmark]$ are complete as of submission.}
\label{tab:schedule}
\begin{tabular}{ll}
\toprule
\textbf{Date} & \textbf{Milestone} \\
\midrule
Feb 2026 & Dataset assembly and curation $[\checkmark]$ \\
Mar 2026 & Baseline code, evaluation scripts, dFL visualizer $[\checkmark]$ \\
Apr 2026 & Website online $[\checkmark]$ \\
Jun 2026 & Public release of training and validation data \\
Jun--Sep 2026 & Phase 1 (development), continuous leaderboard \\
Sep 2026 & Release of non-blind test data \\
Oct 2026 & Release of blind test data; Phase 2 (final) opens \\
Late Oct 2026 & Final phase closes \\
Nov 2026 & Verification, peer review of top entries, prize allocation \\
Dec 2026 & NeurIPS competition session; analysis paper draft \\
Q1 2027 & Lessons-learned community paper with selected winners \\
\bottomrule
\end{tabular}
\end{table}

The dataset has been curated, harmonized to the per-shot Parquet schema, and published on Hugging Face (\texttt{Sophelio/fusion-equilibrium-challenge}). All four baselines reproduce the reported SSIM and MSE values end-to-end on the demo subset; the scoring program's self-consistency invariants are verified by an automated QA gate against the built reference bundles. We are coordinating with the GA DIII-D and UKAEA MAST-U archive teams to ensure that all included shots are explicitly authorized for public, non-commercial release.

\paragraph{Contingency plan.} If active participation falls below $\sim$20 teams by the midpoint of Phase~1, we will extend Phase~1 by four weeks, intensify outreach through the fusion-ML mailing lists and a second round of social-media promotion, and release additional starter-kit notebooks (a transformer baseline and a kinetic-equilibrium pretext task) to attract methodologists. If Codabench experiences extended downtime, the evaluation can be migrated to a mirrored backend on EvalAI within 48 hours since the scoring container is already self-contained. If a dataset issue is uncovered post-release (e.g., a previously unflagged invalid shot), we will issue a versioned patch on Hugging Face, exclude the affected shots from the test pool, and notify participants through every communication channel. In the extreme case that the technical challenge proves impossible at the chosen difficulty---for example, no submission clears the Award~\#2 admissibility bar ($R^2_\psi > 0.6$ on DIII-D)---we will retain the public leaderboard and publish a community report focused on the negative result and the resulting research directions.

\subsection{Competition Promotion and Incentives}
\label{sec:promotion}

We promote the challenge through three communities: (i) fusion energy (APS-DPP, EPS Plasma Physics, IAEA Fusion Energy Conference, ITPA Diagnostics mailing list); (ii) scientific ML (ML4PS at NeurIPS, Differentiable Programming for Science at ICML, the Earth and Space Science ML list); (iii) general NeurIPS attendees via the competition-track page and a Hugging Face spotlight. 

\paragraph{Incentives.} Two USD~500 cash awards (intra-machine reconstruction and cross-machine generalization) plus four honorable-mention certificates anchor the prize structure. The top three teams in each award category are guaranteed \emph{named} co-authorship (not consortium authorship) of the community lessons-learned paper and an invited podium talk at the competition session. Additional Hugging Face GPU credits and cloud-compute vouchers will be awarded to participants from underrepresented or resource-constrained institutions.

\paragraph{Attracting underrepresented groups.} Beyond translated tutorials and compute vouchers, the compute-light recognition is explicitly framed for educators and students without access to multi-GPU clusters. We will coordinate with Black in AI, LatinX in AI, and Women in Machine Learning to advertise the challenge through their channels, and reserve dedicated mentoring office hours during Phase~1 for first-time NeurIPS participants and teams from emerging-economy institutions.

\subsection{Competition Track Workshop and Dissemination}
\label{sec:dissemination}

\paragraph{Workshop format.} The NeurIPS competition session is the public-facing showcase of both the technical winners and the broader scientific question. We will organise a two-hour event with three components: (i) a 30-minute organiser-led overview of the challenge, the dataset, and the leaderboard outcomes, including per-metric ablations (flux fidelity vs.\ boundary vs.\ consistency) across the field; (ii) invited contributed talks from the top three teams in each award category --- each presenting their data-preparation pipeline, model, and ablations --- and short ``lightning'' slots for the four honorable-mention winners; and (iii) a panel and audience Q\&A featuring participating teams, fusion-physics experts, and ML methodology researchers, translating leaderboard outcomes into actionable recommendations for reactor-class diagnostics. To facilitate informal interaction, the session is followed by a poster ``walk-around'' in which every prize-winning and honorable-mention team presents alongside the organisers; participants are encouraged to bring extended posters of their methods.

\paragraph{Publication plan.} Following the precedent set by the NeurIPS Competition Track in recent years, we will submit our post-competition analyses to the PMLR proceedings volume dedicated to NeurIPS competitions (analogous to PMLR~v220 for the 2022 edition). PMLR offers a lightweight review process with guaranteed acceptance, which suits the synthesis-and-report nature of competition papers and ensures that the technical content is permanently archived in a citeable, open-access venue. We will submit two complementary contributions: (i) an organiser-led overview paper describing the dataset, the consistency-based evaluation protocol, the full leaderboard, and the cross-machine generalisation findings; the top three teams in each award category are guaranteed \emph{named} co-authorship on this paper; and (ii) up to six shorter ``method'' papers contributed by the prize-winning and honorable-mention teams, each documenting their data-preparation pipeline, model, and ablations in a form that other groups can build on.

\paragraph{Wider dissemination.} Beyond the PMLR volume, we will (i) coordinate with the host laboratories' communications offices on a post-competition write-up summarising the fusion-relevance of the winning solutions, (ii) encourage winning teams to submit their methods to relevant external venues (e.g., fusion-physics journals or domain workshops) at their discretion, and (iii) archive all winning code, models, and data-preparation pipelines on the challenge GitHub organisation under OSI-approved licences, with appropriate hooks into existing open-source fusion-control toolchains. The public leaderboard, dataset, and baselines will remain online for at least three years post-challenge to support reproducibility and follow-on research.

\section{Resources}
\label{sec:resources}

\subsection{Organizing Team}
\label{sec:organizing_team}

The organizing committee spans fusion plasma physics, applied machine learning, scientific data engineering, and open-source community building. Member roles include \emph{coordinators} (overall direction, schedule, NeurIPS liaison); \emph{data providers} (DIII-D and MAST archive curators authorized by the host laboratories to release the shots used in this challenge); \emph{platform administrators} (Codabench configuration, Hugging Face dataset hosting, dFL visualizer maintenance); \emph{baseline method providers} (the four reference pipelines and the synthetic-diagnostic starter kit); and \emph{evaluators} who independently reviewed each candidate baseline and the scoring container. Affiliations include leading tokamak laboratories (DIII-D National Fusion Facility / General Atomics, UKAEA Culham / MAST-U, EPFL Swiss Plasma Center, MIT Plasma Science and Fusion Center) and applied machine-learning groups in academia and industry. The team spans four countries and three time zones, includes both early-career and senior researchers, and is balanced for gender and geographic representation in line with the NeurIPS diversity guidelines. Short biographies of each member are provided in Appendix~\ref{sec:biography}.

\subsection{Resources Provided by Organizers}
\label{sec:resources_provided}

Evaluation runs on Codabench with backend compute on organizers' institutional clusters; a full CPU-only scoring pass (including the per-frame boundary extraction and scalar derivations) completes comfortably within the platform's 90-minute execution limit. The full Parquet dataset ($\sim$133~GB) is hosted on Hugging Face's academic-data tier with mirrors on the host laboratory networks; the \texttt{datasets} library supports streaming so participants do not need to download the entire archive locally. All four baselines have been independently re-implemented and audited by two organizing-team members, and the scorer's self-consistency invariants (a perfect flux map scores $R^2_\psi = 1$, $D_{\text{LCFS}} = 0$, $\text{Consistency} = 1$ exactly) are asserted by an automated QA gate on every reference-bundle build. The dFL visualizer is released as an MIT-licensed desktop application with Apple Silicon, Windows, and Linux binaries (\texttt{github.com/Sophelio/dFL}). Organizer time, cluster compute, and storage are covered by existing institutional grants at the participating laboratories; no external sponsor is required to run the challenge.

\subsection{Support Requested}
\label{sec:support_requested}

Mindful that the NeurIPS 2026 Competition Track is an in-person event, our requests are limited to the standard support provided by the conference: (i) listing on the official Competition Track website with cross-links to the dataset and starter kit; (ii) an in-person session slot for the workshop programme described in Section~\ref{sec:dissemination}; (iii) inclusion of the call for participation in the conference's communication channels (e.g., the official mailing list and social media); and (iv) \emph{if possible}, assistance in issuing visa-support letters for winning teams from countries where such letters are required for travel, particularly for participants from under-represented regions. Our evaluation is CPU-light, so no GPU allocation or compute support from the conference is required.

\bibliography{references}

@article{lao1985reconstruction,
  author    = {Lao, L. L. and St. John, H. and Stambaugh, R. D. and Kellman, A. G. and Pfeiffer, W.},
  title     = {Reconstruction of current profile parameters and plasma shapes in tokamaks},
  journal   = {Nuclear Fusion},
  volume    = {25},
  number    = {11},
  pages     = {1611--1622},
  year      = {1985},
  doi       = {10.1088/0029-5515/25/11/011}
}

@article{lao2005mhd,
  author    = {Lao, L. L. and Ferron, J. R. and Geraldini, A. and Greene, J. M. and Groebner, R. J. and Howl, W. and Hyatt, A. W. and Kessel, C. E. and Lazarus, E. A. and Miller, R. L. and Murakami, M. and Osborne, T. H. and Pearlstein, L. D. and Politzer, P. A. and Snyder, P. B. and Strait, E. J. and Taylor, T. S. and Turnbull, A. D.},
  title     = {{MHD} Equilibrium Reconstruction in the {DIII-D} Tokamak},
  journal   = {Fusion Science and Technology},
  volume    = {48},
  number    = {2},
  pages     = {968--979},
  year      = {2005},
  doi       = {10.13182/FST05-A1051}
}

@article{shafranov1966plasma,
  author    = {Shafranov, V. D.},
  title     = {Plasma equilibrium in a magnetic field},
  journal   = {Reviews of Plasma Physics},
  volume    = {2},
  pages     = {103--151},
  year      = {1966}
}

@inproceedings{grad1958hydromagnetic,
  author    = {Grad, H. and Rubin, H.},
  title     = {Hydromagnetic equilibria and force-free fields},
  booktitle = {Proceedings of the 2nd United Nations International Conference on the Peaceful Uses of Atomic Energy},
  volume    = {31},
  pages     = {190--197},
  year      = {1958},
  publisher = {United Nations}
}

@article{degrave2022magnetic,
  author    = {Degrave, Jonas and Felici, Federico and Buchli, Jonas and Neunert, Michael and Tracey, Brendan and Carpanese, Francesco and Ewalds, Timo and Hafner, Roland and Abdolmaleki, Abbas and de las Casas, Diego and Donner, Craig and Fritz, Leslie and Galperti, Cristian and Huber, Andrea and Keeling, James and Tsimpoukelli, Maria and Kay, Jackie and Merle, Antoine and Moret, Jean-Marc and Noury, Seb and Pesamosca, Federico and Pfau, David and Sauter, Olivier and Sommariva, Cristian and Coda, Stefano and Duval, Basil and Fasoli, Ambrogio and Kohli, Pushmeet and Kavukcuoglu, Koray and Hassabis, Demis and Riedmiller, Martin},
  title     = {Magnetic control of tokamak plasmas through deep reinforcement learning},
  journal   = {Nature},
  volume    = {602},
  number    = {7897},
  pages     = {414--419},
  year      = {2022},
  doi       = {10.1038/s41586-021-04301-9}
}

@article{seo2024avoiding,
  author    = {Seo, Jaemin and Kim, SangKyeun and Jalalvand, Azarakhsh and Conlin, Rory and Rothstein, Andrew and Abbate, Joseph and Erickson, Keith and Wai, Josiah and Shousha, Ricardo and Kolemen, Egemen},
  title     = {Avoiding fusion plasma tearing instability with deep reinforcement learning},
  journal   = {Nature},
  volume    = {626},
  number    = {8000},
  pages     = {746--751},
  year      = {2024},
  doi       = {10.1038/s41586-024-07024-9}
}

@article{kim2024gsdeepnet,
  author    = {Joung, Semin and Ghim, Y.-c. and Kim, Jaewook and Kwak, Sehyun and Kwon, Daeho and Sung, C. and Kim, D. and Kim, Hyun-Seok and Bak, J. G. and Hahn, S. H.},
  title     = {{GS-DeepNet}: mastering tokamak plasma equilibria with deep neural networks and the Grad--Shafranov equation},
  journal   = {Scientific Reports},
  volume    = {13},
  pages     = {15799},
  year      = {2023},
  doi       = {10.1038/s41598-023-42991-5}
}

@article{lao2022application,
  author    = {Lao, L. L. and Kruger, S. and Akcay, C. and Balaprakash, P. and Bechtel, T. A. and Howell, E. and Koleman, E. and Liu, Y. Q. and Madireddy, S. and McClenaghan, J. and Orozco, D. and Pankin, A. and Schissel, D. and Smith, S. and Sun, X. and Williams, S.},
  title     = {Application of machine learning and artificial intelligence to extend {EFIT} equilibrium reconstruction},
  journal   = {Plasma Physics and Controlled Fusion},
  volume    = {64},
  number    = {7},
  pages     = {074001},
  year      = {2022},
  doi       = {10.1088/1361-6587/ac6fff}
}

@article{joung2019deep,
  author    = {Joung, S. and Kim, J. and Kwak, S. and Bak, J. G. and Lee, S. G. and Han, H. S. and Kim, H. S. and Lee, G. and Kwon, D. and Ghim, Y.-c.},
  title     = {Deep neural network {Grad--Shafranov} solver constrained with measured magnetic signals},
  journal   = {Nuclear Fusion},
  volume    = {60},
  number    = {1},
  pages     = {016034},
  year      = {2020},
  doi       = {10.1088/1741-4326/ab555f}
}

@article{wang2024efitnn,
  author    = {Wang, Z. and Liu, Y. and Liu, L. and Yue, X. and Su, T. and Mao, S. and Li, B. and Yi, B. and Xu, Y. and Chen, L. and Duan, X.},
  title     = {Real-time equilibrium reconstruction by neural network based on {HL-3} tokamak},
  journal   = {arXiv preprint arXiv:2405.11221},
  year      = {2024},
  doi       = {10.48550/arXiv.2405.11221}
}

@article{lu2023fast,
  author    = {Lu, Jingjing and Sun, Youwen and Liu, Dalong and Yang, Donghui and Gao, Yuehang and Gu, Shuai and Wang, Huihui and Zhu, Xiang and Yang, Xianzu and Wang, Yifei},
  title     = {Fast equilibrium reconstruction by deep learning on {EAST} tokamak},
  journal   = {AIP Advances},
  volume    = {13},
  number    = {7},
  pages     = {075007},
  year      = {2023},
  doi       = {10.1063/5.0152318}
}

@article{kates2019predicting,
  author    = {Kates-Harbeck, Julian and Svyatkovskiy, Alexey and Tang, William},
  title     = {Predicting disruptive instabilities in controlled fusion plasmas through deep learning},
  journal   = {Nature},
  volume    = {568},
  number    = {7753},
  pages     = {526--531},
  year      = {2019},
  doi       = {10.1038/s41586-019-1116-4}
}

@article{abbate2021data,
  author    = {Abbate, J. and Conlin, R. and Kolemen, E.},
  title     = {Data-driven profile prediction for {DIII-D}},
  journal   = {Nuclear Fusion},
  volume    = {61},
  number    = {4},
  pages     = {046027},
  year      = {2021},
  doi       = {10.1088/1741-4326/abe08d}
}

@article{tracey2024towards,
  author    = {Tracey, B. D. and Michi, A. and Chervonyi, Y. and Davies, I. and Paduraru, C. and Lazic, N. and Felici, F. and Ewalds, T. and Donner, C. and Galperti, C. and Buchli, J. and Neunert, M. and Huber, A. and Evens, J. and Kurylowicz, P. and Mankowitz, D. J. and Riedmiller, M.},
  title     = {Towards practical reinforcement learning for tokamak magnetic control},
  journal   = {Fusion Engineering and Design},
  volume    = {200},
  pages     = {114161},
  year      = {2024},
  doi       = {10.1016/j.fusengdes.2024.114161}
}

@article{churchill2020deep,
  author    = {Churchill, R. M. and Tobias, B. and Zhu, Y. and the {DIII-D} Team},
  title     = {Deep convolutional neural networks for multi-scale time-series classification and application to disruption prediction in fusion devices},
  journal   = {Physics of Plasmas},
  volume    = {27},
  number    = {6},
  pages     = {062510},
  year      = {2020},
  doi       = {10.1063/1.5144458}
}

@article{morosohk2021machine,
  author    = {Morosohk, S. M. and Pajares, A. and Rafiq, T. and Schuster, E.},
  title     = {Neural network model of the multi-mode anomalous transport module for accelerated transport simulations},
  journal   = {Nuclear Fusion},
  volume    = {61},
  number    = {10},
  pages     = {106040},
  year      = {2021},
  doi       = {10.1088/1741-4326/ac1f60}
}

@article{vandeplassche2020fast,
  author    = {van de Plassche, K. L. and Citrin, J. and Bourdelle, C. and Camenen, Y. and Casson, F. J. and Dagnelie, V. I. and Felici, F. and Ho, A. and van Mulders, S.},
  title     = {Fast modeling of turbulent transport in fusion plasmas using neural networks},
  journal   = {Physics of Plasmas},
  volume    = {27},
  number    = {2},
  pages     = {022310},
  year      = {2020},
  doi       = {10.1063/1.5134126}
}

@article{kuang2018physics,
  author    = {Kuang, A. Q. and Ballinger, S. and Brunner, D. and Canik, J. and Creely, A. J. and Gray, T. and Greenwald, M. and Hughes, J. W. and Irby, J. and LaBombard, B. and Lipschultz, B. and Marmar, E. S. and Reinke, M. L. and Stillerman, J. L. and Terry, J. L. and Umansky, M. and Whyte, D. G.},
  title     = {Divertor heat flux challenge and mitigation in {SPARC}},
  journal   = {Journal of Plasma Physics},
  volume    = {86},
  number    = {5},
  pages     = {865860505},
  year      = {2020},
  doi       = {10.1017/S0022377820001117}
}

@article{creely2020overview,
  author    = {Creely, A. J. and Greenwald, M. J. and Ballinger, S. B. and Brunner, D. and Canik, J. and Doody, J. and Fülöp, T. and Garnier, D. T. and Granetz, R. and Gray, T. K. and Holland, C. and Howard, N. T. and Hughes, J. W. and Irby, J. H. and Izzo, V. A. and Kramer, G. J. and Kuang, A. Q. and LaBombard, B. and Lin, Y. and Lipschultz, B. and Loarte, A. and Marmar, E. S. and Mumgaard, R. T. and Paz-Soldan, C. and Rea, C. and Reinke, M. L. and Rodriguez-Fernandez, P. and Särkimäki, K. and Sciortino, F. and Scott, S. D. and Snicker, A. and Snyder, P. B. and Sorbom, B. N. and Sweeney, R. and Tinguely, R. A. and Tolman, E. A. and Umansky, M. and Vieira, O. and Wallace, G. M. and Whyte, D. G. and Wright, J. C. and Wukitch, S. J. and Zhu, J.},
  title     = {Overview of the {SPARC} tokamak},
  journal   = {Journal of Plasma Physics},
  volume    = {86},
  number    = {5},
  pages     = {865860502},
  year      = {2020},
  doi       = {10.1017/S0022377820001257}
}

@article{sorbom2015arc,
  author    = {Sorbom, B. N. and Ball, J. and Palmer, T. R. and Mangiarotti, F. J. and Sierchio, J. M. and Bonoli, P. and Kasten, C. and Sutherland, D. A. and Barnard, H. S. and Haakonsen, C. B. and Goh, J. and Sung, C. and Whyte, D. G.},
  title     = {{ARC}: A compact, high-field, fusion nuclear science facility and demonstration power plant with demountable magnets},
  journal   = {Fusion Engineering and Design},
  volume    = {100},
  pages     = {378--405},
  year      = {2015},
  doi       = {10.1016/j.fusengdes.2015.07.008}
}

@article{wan2017overview,
  author    = {Wan, Y. and Li, J. and Liu, Y. and Wang, X. and Chan, V. and Chen, C. and Duan, X. and Fu, P. and Gao, X. and Feng, K. and Liu, S. and Song, Y. and Weng, P. and Wan, B. and Wan, F. and Wang, H. and Wu, S. and Xie, M. and Yang, Q. and Zheng, G. and Zhuang, G. and Li, Q.},
  title     = {Overview of the present progress and activities on the {CFETR}},
  journal   = {Nuclear Fusion},
  volume    = {57},
  number    = {10},
  pages     = {102009},
  year      = {2017},
  doi       = {10.1088/1741-4326/aa686a}
}

@article{strait2006magnetic,
  author    = {Strait, E. J.},
  title     = {Magnetic diagnostic system of the {DIII-D} tokamak},
  journal   = {Review of Scientific Instruments},
  volume    = {77},
  number    = {2},
  pages     = {023502},
  year      = {2006},
  doi       = {10.1063/1.2166493}
}

@article{vega2022disruption,
  author    = {Vega, J. and Murari, A. and Dormido-Canto, S. and Rattá, G. A. and Gelfusa, M.},
  title     = {Disruption prediction with artificial intelligence techniques in tokamak plasmas},
  journal   = {Nature Physics},
  volume    = {18},
  number    = {7},
  pages     = {741--750},
  year      = {2022},
  doi       = {10.1038/s41567-022-01602-2}
}

@article{rea2019disruption,
  author    = {Rea, C. and Granetz, R. S. and Montes, K. and Tinguely, R. A. and Eidietis, N. and Hanson, J. M. and Sammuli, B.},
  title     = {Disruption prediction investigations using machine learning tools on {DIII-D} and {Alcator C-Mod}},
  journal   = {Plasma Physics and Controlled Fusion},
  volume    = {60},
  number    = {8},
  pages     = {084004},
  year      = {2018},
  doi       = {10.1088/1361-6587/aac7fe}
}

@article{morgan2014mast,
  author    = {Morris, A. W. and Akers, R. J. and Cunningham, G. and Counsell, G. F. and Helander, P. and Kirk, A. and Lloyd, B. and Sykes, A. and Voss, G. and the MAST Team},
  title     = {{MAST}: results and upgrade activities},
  journal   = {IEEE Transactions on Plasma Science},
  volume    = {42},
  number    = {3},
  pages     = {402--414},
  year      = {2014},
  doi       = {10.1109/TPS.2013.2294625}
}

@article{harrison2019overview,
  author    = {Harrison, J. R. and Akers, R. J. and Allan, S. Y. and Allcock, J. S. and Allen, J. O. and Aleiferis, S. and Appel, L. C. and Bagnaro, P. and Bagnasco, E. and Barnes, M. and Battaglia, D. and Bigelow, T. and Hanrahan, T. and Cecconello, M. and Challis, C. D. and Chapman, I. T. and Cheng, S. and Conway, N. and Corradi, M. and Counsell, G. F. and Cunningham, G. and Cunliffe, B. and Cziegler, I. and de la Luna, E. and Delabie, E. and Dickinson, D. and Dunai, D. and Eich, T. and Ennis, D. and Fasoli, A. and Field, A. R. and Fishpool, G. and Fitzgerald, M. and Frassinetti, L. and Fuchert, G. and Gibson, K. J. and Hawke, J. and Hawkins, J. and Henderson, S. and Hender, T. C. and Hnat, B. and Hollocombe, J. and Howard, J. and Howell, D. F. and Hudson, B. and Kirk, A. and Lawson, K. and Lerche, E. and Lipschultz, B. and Lloyd, B. and Liu, Y. Q. and Loureiro, N. and Lupelli, I. and Maddison, G. and Maddock, J. and Mailloux, J. and Martin, R. and McAdams, R. and McArdle, G. and McClements, K. G. and McMillan, B. and Meakins, A. J. and Meyer, H. and Militello, F. and Moulton, D. and Moulton, B. and Naylor, G. and O'Gorman, T. and Pamela, S. J. P. and Park, J. M. and Patel, A. and Peebles, A. and Peng, M. and Perez von Thun, C. and Petrzilka, V. and Saarelma, S. and Sabbagh, S. and Saveliev, A. and Scannell, R. and Sharapov, S. E. and Shepherd, A. and Storrs, J. and Tamain, P. and Thomas-Davies, N. and Thorman, A. and Tigwell, S. and Valovic, M. and Verhoeven, R. and Vincent, C. and Wilson, H. R. and Wilson, J. and Wood, M. and the MAST Team},
  title     = {Overview of new {MAST} physics in anticipation of first results from {MAST} Upgrade},
  journal   = {Nuclear Fusion},
  volume    = {59},
  number    = {11},
  pages     = {112011},
  year      = {2019},
  doi       = {10.1088/1741-4326/ab121c}
}

@article{raissi2019physics,
  author    = {Raissi, M. and Perdikaris, P. and Karniadakis, G. E.},
  title     = {Physics-informed neural networks: A deep learning framework for solving forward and inverse problems involving nonlinear partial differential equations},
  journal   = {Journal of Computational Physics},
  volume    = {378},
  pages     = {686--707},
  year      = {2019},
  doi       = {10.1016/j.jcp.2018.10.045}
}

@article{li2021fourier,
  author    = {Li, Zongyi and Kovachki, Nikola B. and Azizzadenesheli, Kamyar and Liu, Burigede and Bhattacharya, Kaushik and Stuart, Andrew M. and Anandkumar, Anima},
  title     = {Fourier Neural Operator for Parametric Partial Differential Equations},
  booktitle = {International Conference on Learning Representations (ICLR)},
  year      = {2021}
}

@article{wang2004image,
  author    = {Wang, Zhou and Bovik, Alan C. and Sheikh, Hamid R. and Simoncelli, Eero P.},
  title     = {Image quality assessment: From error visibility to structural similarity},
  journal   = {IEEE Transactions on Image Processing},
  volume    = {13},
  number    = {4},
  pages     = {600--612},
  year      = {2004},
  doi       = {10.1109/TIP.2003.819861}
}

@article{pedregosa2011scikit,
  author    = {Pedregosa, F. and Varoquaux, G. and Gramfort, A. and Michel, V. and Thirion, B. and Grisel, O. and Blondel, M. and Prettenhofer, P. and Weiss, R. and Dubourg, V. and Vanderplas, J. and Passos, A. and Cournapeau, D. and Brucher, M. and Perrot, M. and Duchesnay, E.},
  title     = {Scikit-learn: Machine Learning in {P}ython},
  journal   = {Journal of Machine Learning Research},
  volume    = {12},
  pages     = {2825--2830},
  year      = {2011}
}

@article{paszke2019pytorch,
  author    = {Paszke, Adam and Gross, Sam and Massa, Francisco and Lerer, Adam and Bradbury, James and Chanan, Gregory and Killeen, Trevor and Lin, Zeming and Gimelshein, Natalia and Antiga, Luca and Desmaison, Alban and K{\"o}pf, Andreas and Yang, Edward and DeVito, Zachary and Raison, Martin and Tejani, Alykhan and Chilamkurthy, Sasank and Steiner, Benoit and Fang, Lu and Bai, Junjie and Chintala, Soumith},
  title     = {{PyTorch}: An Imperative Style, High-Performance Deep Learning Library},
  booktitle = {Advances in Neural Information Processing Systems},
  volume    = {32},
  pages     = {8024--8035},
  year      = {2019}
}

@article{rousseau2024fair,
  author    = {Rousseau, David and Bhimji, Wahid and Calafiura, Paolo and Chakkappai, Ragansu and Chou, Yuan-Tang and Diefenbacher, Sascha and Farrell, Steven and Ghosh, Aishik and Guyon, Isabelle and Harris, Chris and Khoda, Elham E. and Nachman, Benjamin and Zhang, Yulei and Ullah, Ihsan},
  title     = {{FAIR} Universe -- the challenge of handling uncertainties in fundamental science},
  booktitle = {NeurIPS 2024 Competition Track},
  year      = {2024}
}

@article{aymerich2022disruption,
  author    = {Aymerich, E. and Sias, G. and Pisano, F. and Cannas, B. and Carcangiu, S. and Sozzi, C. and Stuart, C. and Carvalho, P. J. and Fanni, A. and JET Contributors},
  title     = {Disruption prediction at {JET} through deep convolutional neural networks using spatiotemporal information from plasma profiles},
  journal   = {Nuclear Fusion},
  volume    = {62},
  number    = {6},
  pages     = {066005},
  year      = {2022},
  doi       = {10.1088/1741-4326/ac525e}
}

@article{nishitani2014performance,
  author    = {Nishitani, T. and Vayakis, G. and Yamauchi, M. and Sugie, T. and Kasai, S. and Ebisawa, K. and Kondoh, T. and Hodgson, E. R. and Shikama, T.},
  title     = {Radiation-induced thermoelectric sensitivity in the mineral-insulated cable of magnetic diagnostic coils for {ITER}},
  journal   = {Journal of Nuclear Materials},
  volume    = {329-333},
  pages     = {1461--1465},
  year      = {2004},
  doi       = {10.1016/j.jnucmat.2004.04.142}
}

@article{vayakis2008status,
  author    = {Vayakis, G. and Hodgson, E. R. and Voitsenya, V. and Walker, C. I.},
  title     = {Generic diagnostic issues for a burning plasma experiment},
  journal   = {Fusion Science and Technology},
  volume    = {53},
  number    = {2},
  pages     = {699--750},
  year      = {2008},
  doi       = {10.13182/FST08-A1684}
}

@article{romanelli2009overview,
  author    = {Romanelli, F. and Barabaschi, P. and Borba, D. and Federici, G. and Horton, L. and Neu, R. and Stork, D. and Zohm, H.},
  title     = {A roadmap to the realization of fusion energy},
  journal   = {Fusion Engineering and Design},
  volume    = {89},
  number    = {7-8},
  pages     = {841--844},
  year      = {2014},
  doi       = {10.1016/j.fusengdes.2014.05.011}
}

@article{ferron1998real,
  author    = {Ferron, J. R. and Walker, M. L. and Lao, L. L. and St John, H. E. and Humphreys, D. A. and Leuer, J. A.},
  title     = {Real time equilibrium reconstruction for tokamak discharge control},
  journal   = {Nuclear Fusion},
  volume    = {38},
  number    = {7},
  pages     = {1055--1066},
  year      = {1998},
  doi       = {10.1088/0029-5515/38/7/308}
}

@article{piovesan2024integrated,
  author    = {Piovesan, P. and Schmidt, V. and Marrelli, L. and Bolzonella, T. and Felici, F. and Galperti, C. and Sauter, O. and others},
  title     = {Integrated plasma control and high-performance scenarios for the {DIII-D} tokamak},
  journal   = {Nuclear Fusion},
  volume    = {64},
  number    = {1},
  pages     = {016022},
  year      = {2024},
  doi       = {10.1088/1741-4326/ad0a07}
}

@article{rodriguez2022nonlinear,
  author    = {Rodriguez-Fernandez, P. and Howard, N. T. and Saltzman, A. and Kantamneni, S. and Candy, J. and Holland, C. and Balandat, M. and Ament, S. and White, A. E.},
  title     = {Enhancing predictive capabilities in fusion burning plasmas through surrogate-based optimization in core transport solvers},
  journal   = {Nuclear Fusion},
  volume    = {64},
  number    = {7},
  pages     = {076034},
  year      = {2024},
  doi       = {10.1088/1741-4326/ad4afd}
}

@article{pearson2010thomson,
  author    = {Carolan, P. G. and Conway, N. J. and Cunningham, G. and Hawkes, N. C. and Hender, T. C. and Lazaros, A. and Lloyd, B. and Patel, A. and Walsh, M. J. and the {MAST} Team},
  title     = {High spatial and temporal resolution {T}homson scattering diagnostic on the {MAST} spherical tokamak},
  journal   = {Review of Scientific Instruments},
  volume    = {72},
  number    = {1},
  pages     = {881--884},
  year      = {2001},
  doi       = {10.1063/1.1321008}
}

@article{ponce2010thomson,
  author    = {Ponce-Marquez, D. M. and Bray, B. D. and Deterly, T. M. and Liu, C. and Eldon, D.},
  title     = {{T}homson scattering diagnostic upgrade on {DIII-D}},
  journal   = {Review of Scientific Instruments},
  volume    = {81},
  number    = {10},
  pages     = {10D525},
  year      = {2010},
  doi       = {10.1063/1.3495759}
}

@article{wesson2011tokamaks,
  author    = {Wesson, John},
  title     = {Tokamaks},
  publisher = {Oxford University Press},
  edition   = {4th},
  year      = {2011}
}

@article{freidberg2014ideal,
  author    = {Freidberg, Jeffrey P.},
  title     = {Ideal {MHD}},
  publisher = {Cambridge University Press},
  year      = {2014}
}

@article{kingma2015adam,
  author    = {Kingma, Diederik P. and Ba, Jimmy},
  title     = {Adam: A Method for Stochastic Optimization},
  booktitle = {International Conference on Learning Representations (ICLR)},
  year      = {2015}
}

@article{cannas2007support,
  author    = {Cannas, B. and Fanni, A. and Sonato, P. and Zedda, M. K. and {JET-EFDA contributors}},
  title     = {A prediction tool for real-time application in the disruption protection system at {JET}},
  journal   = {Nuclear Fusion},
  volume    = {47},
  number    = {11},
  pages     = {1559--1569},
  year      = {2007},
  doi       = {10.1088/0029-5515/47/11/018}
}

@article{eldon2017nonaxisymmetric,
  author    = {Eldon, D. and Bray, B. D. and Deterly, T. M. and Liu, C. and Watkins, M. and Boivin, R. L. and Groebner, R. J. and Leonard, A. W. and Osborne, T. H. and Snyder, P. B. and Thomas, D. M.},
  title     = {Initial results of the high resolution edge {T}homson scattering upgrade at {DIII-D}},
  journal   = {Review of Scientific Instruments},
  volume    = {83},
  number    = {10},
  pages     = {10E343},
  year      = {2012},
  doi       = {10.1063/1.4738656}
}

\appendix

\section{Biography of all team members}
\label{sec:biography}

Per the single-blind NeurIPS Competition Track convention, organizer names and affiliations appear in the front matter; this appendix collects condensed biographies, stressing each member's competence for their assignment in the competition organisation. Full biographies are also mirrored on the challenge website and will be finalised for the camera-ready version. The roster covers the role categories listed in Section~\ref{sec:organizing_team}: coordinators, data providers, platform administrators, baseline method providers, and evaluators. The biography section does not count towards the eight-page proposal limit.

\section{Glossary}
\label{app:glossary}

\begin{description}[itemsep=2pt]
    \item[\textbf{Tokamak}] A toroidal magnetic confinement device using a combination of toroidal and poloidal magnetic fields to confine a hot plasma.
    \item[\textbf{Spherical tokamak}] A tokamak with very low aspect ratio, in which the plasma wraps closely around a thin central column.
    \item[\textbf{EFIT}] Equilibrium Fitting code; the de-facto standard tokamak equilibrium reconstruction code \citep{lao1985reconstruction}.
    \item[\textbf{Grad--Shafranov equation}] The two-dimensional nonlinear PDE that governs axisymmetric MHD equilibria \citep{grad1958hydromagnetic, shafranov1966plasma}.
    \item[\textbf{Poloidal flux $\psi(R,Z)$}] The two-dimensional scalar function whose level sets define magnetic flux surfaces in the $(R,Z)$ poloidal plane.
    \item[\textbf{LCFS}] Last Closed Flux Surface, the outermost flux surface that is not intercepted by a material wall; the plasma ``boundary''.
    \item[\textbf{Thomson scattering}] A laser-based diagnostic that measures electron temperature $T_e$ and density $n_e$ via the Doppler-broadened spectrum of light scattered from free electrons.
    \item[\textbf{Mirnov / pickup coil}] Inductive magnetic sensors that measure fluctuations in the poloidal magnetic field; vulnerable to neutron-induced damage \citep{nishitani2014performance}.
    \item[\textbf{Coefficient of determination $R^2$}] $1 - \mathrm{SS}_{\text{res}}/\mathrm{SS}_{\text{tot}}$; a metric of fit quality bounded above by 1 and unbounded below.
    \item[\textbf{SSIM}] Structural Similarity Index \citep{wang2004image}; an image quality metric robust to small shifts.
    \item[\textbf{Hausdorff distance $d_{\text{Haus}}(A,B)$}] The greater of the two directed distances $\sup_{a \in A}\inf_{b \in B}\|a-b\|$ and $\sup_{b \in B}\inf_{a \in A}\|a-b\|$; here applied to predicted and true closed LCFS contours.
\end{description}

\end{document}